\documentclass[journal=jacsat,manuscript=article]{achemso}

\usepackage[version=3]{mhchem} 
\usepackage[dvipsnames]{xcolor}
\usepackage{mathtools}
\usepackage{siunitx}
\usepackage{hyperref}
\hypersetup{hidelinks}
\usepackage[normalem]{ulem}
\usepackage{soul}
\usepackage{caption}
\usepackage{subcaption}
\usepackage{booktabs}
\usepackage{tabularx}
\usepackage{array}
\usepackage{color}
\newcolumntype{Y}{>{\raggedright\arraybackslash}X}
\newcommand{\doublerule}{\toprule[0.35pt]\specialrule{0.35pt}{1pt}{0pt}}
\newcommand{\sectionrow}[1]{\multicolumn{4}{@{}l}{\textbf{#1}}\\}
\newcommand{\subsectionrow}[1]{\multicolumn{4}{@{}l}{\hspace{1em}\textit{#1}}\\}

\DeclareSIUnit\angstrom{\text{\AA}}
\DeclareSIUnit\bar{bar}
\DeclareSIUnit{\calorie}{cal}
\DeclareSIUnit\rydberg{Ry}
\DeclareSIUnit\molar{M}

\author{Thorben Eggert}
\affiliation[Fritz-Haber-Institut der Max-Planck-Gesellschaft]
{Theory Department, Fritz-Haber-Institut der Max-Planck-Gesellschaft, Berlin, Germany}

\author{Lang Li}
\affiliation[Fritz-Haber-Institut der Max-Planck-Gesellschaft]
{Theory Department, Fritz-Haber-Institut der Max-Planck-Gesellschaft, Berlin, Germany}

\author{Yair Litman}
\affiliation[MPIP]
{Max Planck Institute for Polymer Research, Ackermannweg 10, 55128 Mainz, Germany}

\author{Karsten Reuter}
\affiliation[Fritz-Haber-Institut der Max-Planck-Gesellschaft]
{Theory Department, Fritz-Haber-Institut der Max-Planck-Gesellschaft, Berlin, Germany}

\author{Nicolas G. H\"ormann}
\affiliation[Fritz-Haber-Institut der Max-Planck-Gesellschaft]
{Theory Department, Fritz-Haber-Institut der Max-Planck-Gesellschaft, Berlin, Germany}
\email{hoermann@fhi.mpg.de}

\author{Clotilde S. Cucinotta}
\affiliation[Fritz-Haber-Institut der Max-Planck-Gesellschaft]
{Theory Department, Fritz-Haber-Institut der Max-Planck-Gesellschaft, Berlin, Germany}
\alsoaffiliation[Imperial College London]
{Department of Chemistry, Faculty of Natural Sciences, Imperial College London, London, UK}
\email{c.cucinotta@imperial.ac.uk}

\title{Disentangling Surface Charge and Electrolyte Effects on Interfacial Water at Electrified Pt(111)}

\keywords{electrified interface, water structure, ab initio molecular dynamics, hydrogen-bonding, Pt(111), electrochemistry}

\begin{document}

\begin{abstract}
The structure of interfacial water at electrified metal electrodes is known to affect electrocatalysis, and experiments highlight its sensitivity to electrolyte ion identity and applied bias. To disentangle generic charge-controlled structural changes from ion-specific effects, we compare \textit{ab initio} molecular dynamics water structures at electrified Pt(111) generated with two very different biasing schemes: an explicit ion imbalance in the double layer and homogeneously distributed partially charged hydrogen atoms. Despite these distinct counter-charge representations, both approaches yield a consistent average response of the first water bilayer, in particular of the chemisorbed first layer, when compared on a common surface-charge scale. Method-dependent differences, in particular those associated with explicit electrolyte ions, become apparent only in more local structural descriptors. The hydrogen-bond topology reveals charge-dependent chain-to-ring rearrangements, with explicit ions enhancing ring populations near the potential of zero charge. Layer-resolved vibrational density of states (VDOS) assigns the strongest O--H stretching perturbation to chemisorbed first-layer water and identifies high-frequency signatures of ion-coordinated water, while computed vibrational sum-frequency generation (VSFG) spectra show that the physisorbed, electrolyte-facing region is particularly sensitive to the counter-charge representation. These results suggest that surface charge controls the average structural response, whereas electrolyte ions and their solvation shells become visible only when locally refined descriptors are investigated or probed, e.g. the H-bond network topology and vibrational fingerprints of the interfacial bilayer.
\end{abstract}

\cleardoublepage

\section{Introduction}

Electrified metal--water interfaces are fundamental to electrocatalytic applications, such as fuel cells and electrolyzers. In these applications, water is far more than an inert medium: it can act as a thermodynamic reservoir for reacting chemical species, e.g., H$^+$ or OH$^-$, or compete with other adsorbates to bind to the electrode.\cite{Heenen2020Solvation, Eggert2023} Interfacial water can drastically alter the electrostatic potential via its strong dipole and charge transfer.\cite{Le2018structure, Li_ACSElectro_2024} Therefore, understanding the structure of water at metal--water interfaces establishes a stepping stone for many other applications at electrified interfaces, including the prediction of reaction barriers and the screening of adsorption energies.

Describing the structure of interfacial water is challenging: even two-dimensional water structures exhibit a large variety of different structural motifs, such as different-sized ring structures,\cite{roman2016polymorphism, Kapil2022, DavilaLopez2021} and their ordering naturally depends on the substrate and its interaction strength with water.\cite{Le2018structure,Bellarosa2016,gim2019structure}

At electrified interfaces, the applied charge becomes an additional control variable: it reshapes the already substrate-sensitive structure and couples the orientational and hydrogen-bond ordering of interfacial water to the electrode state. Capturing this charge response, however, remains a challenge for first-principles modeling, since \textit{ab initio} molecular dynamics (AIMD) simulations are typically conducted in canonical ensembles of fixed overall charge rather than at a fixed applied potential.\cite{Le2020change, Sakong2020c, Khatib2021nanoscale, Darby2022, Buraschi2024, Ahart2024, Raffone2025, Li_ACSElectro_2024, Li2025b, Darby2026} Different methods exist to introduce a charge bias or electrochemical environment under these boundary conditions, e.g., via explicit counterions,\cite{Le2020change} H$^+$ or OH$^-$,\cite{Sakong2020c} an imbalance of counter and co-ions,\cite{Khatib2021nanoscale} a homogeneous background charge,\cite{Li2025b} partially charged hydrogen atoms,\cite{Li_ACSElectro_2024} or continuum/implicit-solvent and implicit-electrolyte treatments.\cite{Mathew2014VASPsol,Sakong2015ImplicitPt,Mathew2019ImplicitElectrolyte} Open-boundary and potential-control approaches provide complementary routes to model biased electrochemical interfaces more directly.\cite{Buraschi2024,Ahart2024} A central question is whether these methods produce comparable results or instead highlight certain specific aspects of the electrified interface.

Because electrode charging and electrolyte ion accumulation are intrinsically coupled in real double layers, a central challenge is to disentangle which changes in interfacial water structure are caused by surface charge itself and which require the explicit presence of electrolyte ions. In this study, we systematically compare two efficient but fundamentally different methods for introducing surface charge within canonical (constant-charge) \textit{ab initio} molecular dynamics: the ion-imbalance method (IIM)\cite{Khatib2021nanoscale} and the partially charged hydrogen (PCH) method.\cite{Li_ACSElectro_2024} Unlike grand-canonical or potential-control approaches,\cite{Buraschi2024,Ahart2024} both methods fix the total charge in the simulation cell rather than the electrode potential, making the applied surface charge a natural comparison variable. Our focus is on analyzing the structure of interfacial water as a function of this surface excess charge at pristine Pt(111), using the comparison to separate generic, charge-controlled response properties from setup-specific contributions such as those related to explicit electrolyte ions.

We first establish the applied surface charge as the common comparison variable by relating the nominal charge imposed in each biasing scheme to the Bader-partitioned charge of the electrode-water interfacial region. We then evaluate the average density and orientational response of the first water bilayer and show that the PCH and IIM descriptions agree on this common surface-charge scale. Beyond average profiles, we analyze dominant orientational regions and hydrogen-bond network topology, quantified through donor/acceptor balance and ring statistics, as complementary descriptors of the charge-dependent local interfacial structure. We observe a general chain-to-ring transition when moving towards positive bias and H-bond network modifications due to the presence of explicit ions and their solvation shells mainly located at the electrolyte-facing side of the bilayer. Finally, we connect these structural descriptors to experimentally accessible vibrational observables via the layer-resolved vibrational density of states (VDOS), which reveals O--H stretching shifts in different interfacial water environments, while vibrational sum-frequency generation (VSFG) adds orientational sensitivity to the physisorbed, electrolyte-facing part of the interface.

\section{Computational Methods}

All simulations were performed using Born--Oppenheimer \textit{ab initio} molecular dynamics (AIMD) as implemented in the Quickstep module of the CP2K code.\cite{CP2K} The electronic structure was obtained at the level of the Perdew--Burke--Ernzerhof (PBE) generalized-gradient approximation\cite{PBE} with Goedecker--Teter--Hutter (GTH) pseudopotentials.\cite{GTH_Goedecker1996,GTH_Hartwigsen1998} Brillouin-zone sampling was restricted to the $\Gamma$ point.

Two cell sizes were employed to model the Pt(111)/water interface. The larger cell, following the IIM setup of Khatib \textit{et al.}~for Pt(111)/water,\cite{Khatib2021nanoscale} uses a $(\sqrt{3} \times 5)$ surface unit cell (in-plane dimensions $a = \SI{14.568}{\angstrom}$, $b = \SI{19.626}{\angstrom}$; cross-sectional area $\SI{286}{\angstrom\squared}$) and was used for both IIM and PCH simulations. It contains two four-layer Pt slabs with 42~Pt atoms per layer in each slab, i.e., 336~Pt atoms in total, and 250~water molecules filling the electrolyte region (${\approx}\,1100$ nuclei, depending on the ion composition). The cell parameter along the surface normal, $c$, ranges from \SIrange{35.9}{39.6}{\angstrom} depending on the electrolyte composition, corresponding to an electrode separation of $d \approx \SI{28}{\angstrom}$ (${\approx}\,8$ Debye screening lengths at ${\approx}\,\SI{2}{\molar}$ NaCl). The coordinates of the central Pt layers were constrained to their bulk positions. For these IIM cells, valence triple-zeta (TZV) basis sets were used for Pt and TZV with polarization (TZVP) for O, H, Na, and Cl, together with a plane-wave cutoff of \SI{300}{\rydberg}. The equations of motion were integrated with a \SI{0.7}{\femto\second} time step in the NVT ensemble using a Langevin thermostat at \SI{340}{\kelvin}.\cite{Khatib2021nanoscale} Each charged system was equilibrated for ${{\approx}}\,\SI{10}{\pico\second}$ and subsequently sampled for ${{\approx}}\,\SI{50}{\pico\second}$. For the IIM approach, the electrolyte consisted of varying numbers of Na$^+$ and Cl$^-$ ions ($12$Na:$10$Cl, $10$Na:$10$Cl, $10$Na:$12$Cl), with the charge imbalance between cation and anion counts determining the nominal surface charge.\cite{Khatib2021nanoscale}

The smaller $(\sqrt{3} \times 3)$ cell ($\SI{8.443}{\angstrom} \times \SI{9.749}{\angstrom} \times \SI{27.894}{\angstrom}$; surface area $\SI{82.3}{\angstrom\squared}$) was used for PCH simulations only, allowing exploration of a wider charge range.\cite{Li2025b} In these runs, Grimme D3 dispersion corrections\cite{GrimmeD3} were included and the plane-wave cutoff was raised to \SI{400}{\rydberg}; GTH pseudopotentials with valence configurations O(q6), H(q1), and Pt(q10) were employed. For the PCH method, the core charges of all hydrogen atoms in water were uniformly modified by $\Delta q_{\mathrm{H}} = eN_e/N_{\mathrm{H}}$ to transfer fractional electrons between the electrolyte and the electrode, thereby avoiding ion-specific effects.\cite{Li_ACSElectro_2024} Production trajectories of ${\approx}\,\SI{30}{\pico\second}$ were collected after equilibration. 

The definition of water layers used throughout this work is based on the bimodal density distribution of water at the Pt(111) surface. The first layer (chemisorbed water) extends up to \SI{2.7}{\angstrom} from the surface for both setups; the second layer (physisorbed water) extends up to \SI{4.5}{\angstrom} (PCH) or \SI{4.8}{\angstrom} (IIM). Water molecules beyond these boundaries are classified as bulk water.

Hydrogen bonds were identified using a geometric criterion\cite{Natarajan2016} ($d_{\mathrm{OO}} \leq \SI{3.5}{\angstrom}$ and $\alpha_{\mathrm{OOH}} \leq \SI{30}{\degree}$). Ring structures in the hydrogen-bond network were identified by constructing an undirected graph from the hydrogen bonds and finding all elementary cycles (containing no smaller rings) using a custom script based on the NetworkX package,\cite{Hagberg2008} with periodic boundary conditions enforced.

The VDOS was calculated from the Fourier transform of hydrogen--hydrogen velocity autocorrelation functions, spatially resolved by the position of the corresponding oxygen atom, following the approach of Le \textit{et al.}\cite{Le2018Theoretical}. VSFG spectra were computed using the surface-specific velocity--velocity correlation function (ss-VVCF) formalism\cite{ssVVCF}.

Unless stated otherwise, applied charges in the structural analysis are reported as excess elementary charge per exposed surface Pt atom ($e/\mathrm{Pt}$); negative (positive) values correspond to electron accumulation on (depletion from) the electrode, i.e., to cathodic (anodic) conditions. For the Bader analysis, total simulation-cell charges and normalized surface charges are used as indicated in Fig.~\ref{fig:bader}; the SI table reports total layer charges in units of $|e|$ per simulation cell.

\section{The Charging Mechanism}

Before turning to the individual analyses, we note that Table~\ref{tab:water_structure_summary} at the end of the Vibrational Analysis section compiles all charge- and electrolyte-dependent trends in structure, hydrogen-bond topology, and vibrational properties resolved by the two methods, and may serve as a compact reference throughout.

\begin{figure}
     \centering
    \includegraphics{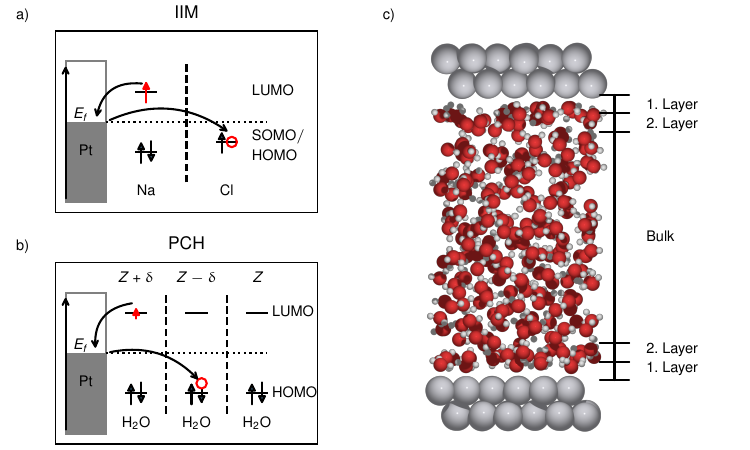}
    \caption{Schematic of the charging mechanisms of the two employed methods: (a) ion-imbalance method (IIM) and (b) partially charged hydrogen (PCH) atoms. The dotted lines mark the Fermi level of the platinum electrode, and simplified energy levels for the species involved in charging are shown (IIM: Na/Cl; PCH: $\mathrm{H_2O}$). The black arrows indicate the occupation of the molecular orbitals, while the red circles (electron shortage) and red arrows (electron surplus) represent initial conditions before the electrons redistribute to more stable states (curved arrows). For the IIM, this transfer involves (ideally) integer numbers of electrons, while in the PCH method, fractional electrons are transferred to all water molecules to compensate for the changed core charges. (c) Example snapshot of the Pt(111)/$\mathrm{H_2O}$ interface at the potential of zero charge. The different layers are labeled on the right side. The upper bounds for the different layers in the PCH (IIM) are: \SI{2.7}{\angstrom} (\SI{2.7}{\angstrom}) for the first layer and \SI{4.5}{\angstrom} (\SI{4.8}{\angstrom}) for the second layer. Water molecules beyond these boundaries are classified as bulk water.}
    \label{fig:setup}
\end{figure}

A graphical depiction of the surface charging mechanisms of both methods is schematically shown in Figs.~\ref{fig:setup}a and \ref{fig:setup}b, together with a snapshot of the system illustrating the dimensions of the setup in Fig.~\ref{fig:setup}c. Curved black arrows indicate electron transfer between the electrode and the electrolyte.
In the IIM, an unequal number of ions in the electrolyte solution generates a charge imbalance. For example, a cation excess results in the transfer of approximately one electron from the electrolyte to the electrode, whereas an anion excess transfers one electron in the opposite direction (see Fig.~\ref{fig:setup}a). We refer to this charge as the nominal surface charge of the electrode.
In the PCH method, no electrolyte ions are added; instead, the core charge of all hydrogen atoms is modified. This drives electron transfer between the water and the electrode. Increasing the core charge partially populates the antibonding LUMO of water, and these excess electrons are transferred to the electrode. Conversely, decreasing the core charge partially depletes the HOMO, leaving a deficit that is compensated by electrons from the metal (see Fig.~\ref{fig:setup}b).
Thus, the former leads to a negative nominal surface charge of the electrode, analogous to a cation excess in the IIM, while the latter leads to a positive nominal surface charge.

\subsection{Charge partitioning and comparison scale}
\label{sec:charging}

\begin{figure}
     \centering
    \includegraphics{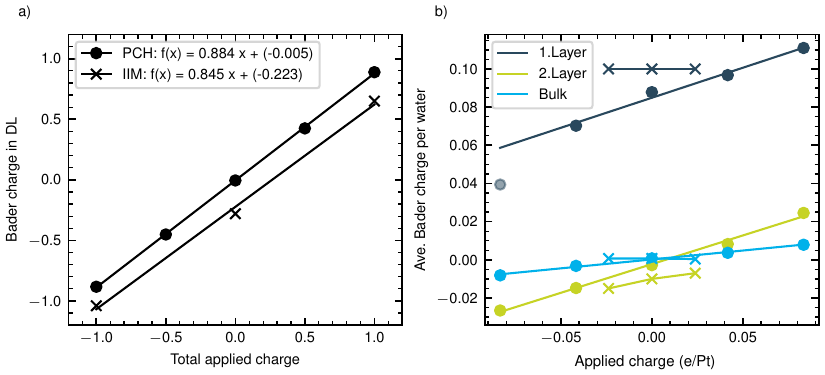}
    \caption{Comparison between applied (nominal) charges and Bader charges. (a) Total nominal charge versus Bader charge in the double layer as defined in Ref.~\citenum{Khatib2021nanoscale} (including contributions from Pt(111) and the first and second water layers). The two setups are represented by different markers and the solid lines are linear fits (see figure legend). (b) Bader charges in the different layers as a function of the nominal surface charge normalized by the number of exposed surface Pt atoms, $e/\mathrm{Pt}$. Panel (a) compares total nominal simulation-cell charge with the total Bader charge of the double-layer region, whereas panel (b) uses the normalized charge scale employed for the structural trends. For both plots, the smaller cells from Ref.~\citenum{Li_ACSElectro_2024} were used for PCH. Data for IIM were taken from Ref.~\citenum{Khatib2021nanoscale}.}
    \label{fig:bader}
\end{figure}

The two charging methods define the applied electrode charge in different ways: in IIM through the ion-count imbalance, and in PCH through the fractional charge introduced by modified hydrogen cores. To assess whether these nominal charges can nevertheless be used on a common scale for the bias, we compare them to the actual amount of charge located in the interfacial region via Bader charge analysis.

\textbf{Bader charge analysis.}
Figure~\ref{fig:bader}a shows that the nominal applied charge and the Bader charge of the double-layer region (electrode and interfacial water bilayer) are linearly correlated for both methods, with slopes close to unity. The nominal surface charge is therefore a practical descriptor for comparing charge-dependent trends across the two setups, even though it is not identical to the Bader-partitioned charge. The slight offset in the IIM fit likely reflects the finite Bader charge assigned to solvated ions, the chosen double-layer partitioning, and differences in electrolyte and cell setup. 

The layer-resolved Bader analysis further shows that chemisorbed first-layer water carries a finite net charge of order $0.10\,|e|$ per molecule. This value is nearly constant in IIM and somewhat more charge-dependent in PCH, but remains of the same order in both approaches. 

The partial charge of chemisorbed water is consistent with the previously reported electron spillover into the first-layer water,\cite{Li2025b, Li2026} and the understanding that chemisorbed water lowers the metal work function through charge transfer to the substrate\cite{Le2018structure,Le2020change}. To put it into perspective: the effective charge of a solvated ion (${\sim}\,0.8\,|e|$, as noted above) is roughly eight times larger than the charge per water molecule. However, because there are typically 5--8 first-layer water molecules per ion in the simulation cell,\cite{Khatib2021nanoscale} the partial charge carried by interfacial water constitutes a significant part of the double-layer polarization. Notably, this effect appears to be specific to Pt: comparative simulations on Ag(111) yield a spillover of only $\sim$0.01$\,|e|$ per first-layer water molecule,\cite{Khatib2021nanoscale} consistent with the weaker chemisorption and lower work function of silver relative to platinum.\cite{Surendralal2021}

\textbf{Choice of the comparison scale.}
Overall, it seems valid to assume that average bias-charge-induced interfacial field strengths are close to identical for both methods when using the nominal charge as scale for the bias strength, which is why we will use it in the subsequent analysis. Unless explicitly stated otherwise, the applied charge is reported as charge per surface Pt atom, $e/\mathrm{Pt}$; figures using total excess electrons per periodic cell are labeled as such.
As we demonstrate below, this charge scale enables meaningful theory--theory comparisons of interfacial structure across the two biasing schemes. Comparisons at a fixed electrode potential are less direct here, because the two computational setups use different potential references and span very different reported capacitance ranges (\SIrange{20}{110}{\micro\farad\per\square\centi\metre} for PCH~\cite{Li_ACSElectro_2024} and \SI{8.29}{\micro\farad\per\square\centi\metre} for the IIM setup~\cite{Khatib2021nanoscale}). Consequently, the same nominal potential difference would correspond to different interfacial field strengths. Since the local structural response is governed primarily by these interfacial fields, the nominal surface charge provides the more transparent comparison variable for the present analysis.

\section{Density and Orientation of Interfacial Water}
\subsection{Average water response}

\begin{figure}
     \centering
    \includegraphics{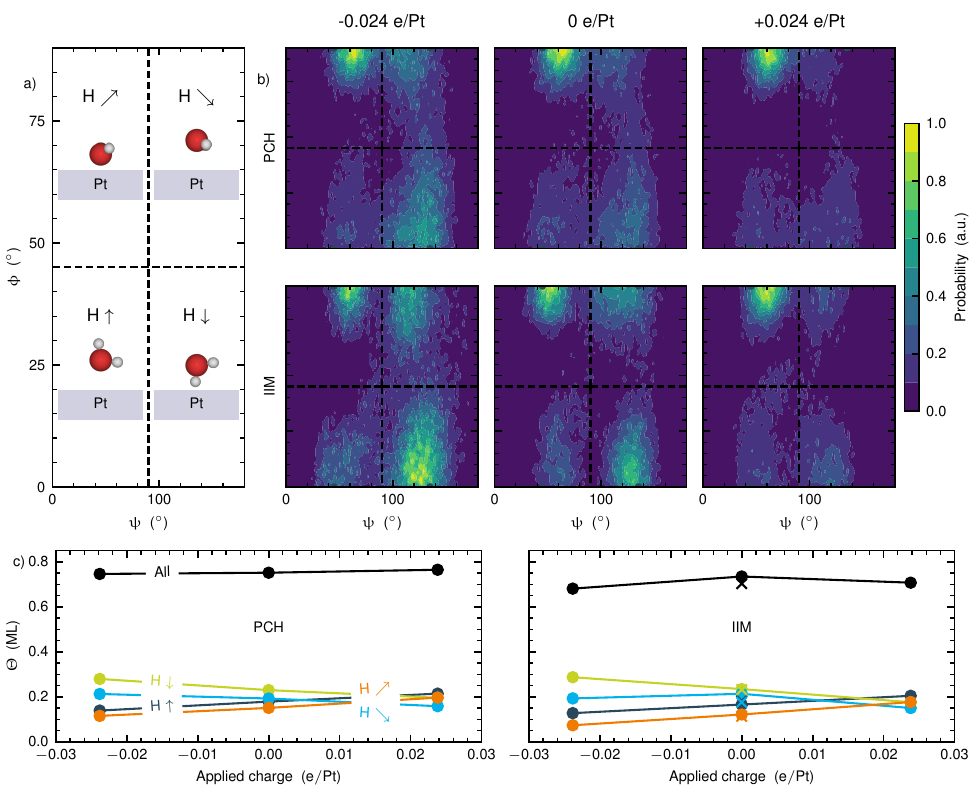}
    \caption{Orientation of interfacial water under applied charge. (a) The four main orientations identified by the cartwheel and propeller descriptors and the tessellation (dashed lines) of the ($\psi$,$\phi$)-plane. (b) ($\psi,\phi$) heat maps of interfacial water for different applied charges; from left to right, the columns correspond to $-0.024$, $0$, and $+0.024\,e/\mathrm{Pt}$. In the upper panels, partially charged hydrogen (PCH) atoms were used for charging, while in the lower panels, the ion-imbalance method (IIM) was used. The color coding represents the relative probabilities. (c) Coverage of each orientation as a function of the applied charge for both charging methods.}
    \label{fig:orientation}
\end{figure}

The Pt(111)/water interface is one of the most fundamental and thoroughly investigated systems in electrochemistry due to its technological relevance, e.g., in the hydrogen evolution reaction. Consequently, this interface has been the focus of many computational AIMD and potential-controlled interface studies.\cite{Bouzid2018, Le2018structure, Le2018Theoretical, Le2020change, Sakong2020c, Khatib2021nanoscale, Darby2022, Buraschi2024, Tang2024, Huang2023, Kristoffersen2018, Zhu2024, Li2022Hydrogen, Heenen2020Solvation, Raffone2025, Darby2026}

These studies revealed a bimodal density distribution of interfacial water within a distance of $d \sim \SIrange{2}{4.5}{\angstrom}$ from the Pt surface, consisting of chemisorbed ($d \sim \SIrange{2}{2.5}{\angstrom}$, 1st layer) and physisorbed water ($d \sim \SIrange{2.5}{4.5}{\angstrom}$, 2nd layer). The tightly bound chemisorbed water molecules are oriented almost flat (cf.\ $\mathrm{H \nearrow}$ in Fig.~\ref{fig:orientation}a), and their rotation is hindered. In contrast, physisorbed water molecules are less strongly bound and can rotate rather freely.

The preference of water for chemisorption or physisorption changes with applied potential\cite{Buraschi2024} or, in constant-charge simulations, with applied surface charge.\cite{Bouzid2018, Le2020change, Sakong2020c, Khatib2021nanoscale, Li_ACSElectro_2024} More positive electrode charging promotes chemisorption, while physisorption is favored at more negative electrode charge. In addition, physisorbed water molecules reorient and align their dipole with the electric field.\cite{Bouzid2018, Le2020change, Sakong2020c, Goldsmith2021, Tang2024} This generic reorientation is captured by simple classical force fields\cite{Scalfi2020, Schlaich2019a, Dewan2014, Willard2009} and DFT-based interface models,\cite{Buraschi2024, Hinsch2023} whereas the chemisorption behavior is not reproduced by classical force fields. At the potential of zero charge (PZC), both the orientational distribution of interfacial water and its response to charging are closely similar for the two methods when compared at the same excess surface charge (see also Fig.~S1 in the Supporting Information (SI)). The extended PCH series is therefore used here to map charge-dependent trends over a wider range, while the IIM data provide the complementary information that is inaccessible in ion-free PCH simulations: the local effects of explicit electrolyte ions, their solvation shells, and electrolyte screening on otherwise charge-controlled interfacial structures.

\subsection{Orientation of interfacial water}\label{sec:orientation}

Beyond the ensemble-averaged profiles, we resolve the instantaneous orientation of each interfacial water molecule into discrete orientational classes. For this purpose, we used the cartwheel ($\psi$) and propeller ($\phi$) angles of each water molecule, as defined in Ref.~\citenum{Clabaut2020}, as orthogonal descriptors to characterize their orientation within the first water bilayer (chemisorbed and physisorbed water). The former is the angle between the surface normal and the water bisector, while the latter describes the rotation around the bisector (for $\phi=0$, the molecular plane of $\mathrm{H_2O}$ and the (bisector, $z$-axis) plane are parallel). Plotting the 2D density profile as a function of these two orientational descriptors reveals four separated peaks corresponding to distinct interfacial water orientations (cf.\ Fig.~\ref{fig:orientation}a/b).

Each of these orientations is well-documented in the prior Pt(111)/water AIMD literature: the chemisorbed $\mathrm{H \nearrow}$ configuration, with the HOH plane slightly tilted away from the metal,\cite{Bouzid2018, Le2018structure} and three physisorbed configurations with one O--H pointing toward the bulk ($\mathrm{H \uparrow}$),\cite{Le2020change} one O--H pointing toward the metal ($\mathrm{H \downarrow}$),\cite{Sakong2020c} and both O--H bonds lying approximately parallel to the surface ($\mathrm{H \searrow}$).\cite{Clabaut2020} Separate plots for the first and second layers are shown in the SI. Each interfacial water molecule is assigned to one of these orientations based on its $(\psi,\phi)$ coordinates. Changes in the interfacial water structure as a function of applied charge are then captured as gradual changes in the relative populations of these orientations, while the total number of molecules in the first water bilayer remains largely invariant. This invariance refers to the combined first and second layers; charging redistributes molecules between chemisorbed and physisorbed regimes rather than changing the overall bilayer population (Fig.~\ref{fig:orientation}c). The coverage of each orientation changes approximately linearly with the applied charge, reproducing trends reported in the literature.\cite{Bouzid2018, Le2020change, Sakong2020c} We find good qualitative agreement between both charging methods in this analysis, including the magnitudes of the orientation populations and their changes with the nominal surface charge of the electrode.

\section{Hydrogen-Bond Network Topology}

\begin{figure}
     \centering
    \includegraphics{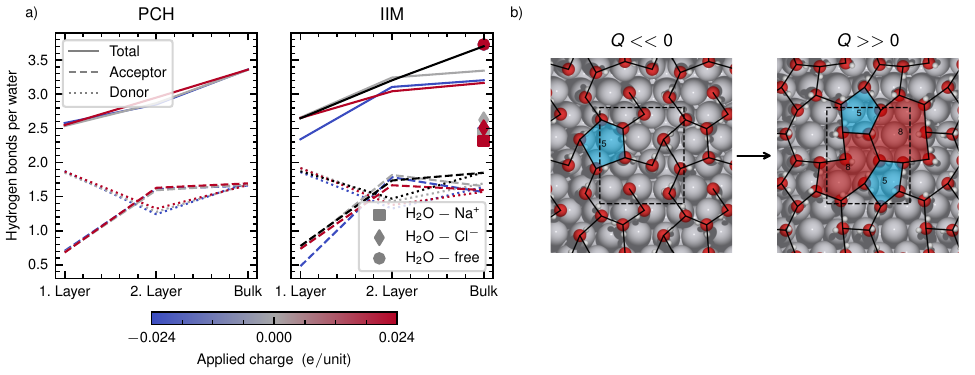}
    \caption{Number of hydrogen-bond donors and acceptors per water molecule in the different layers at different applied charges. (a) Comparison between partially charged hydrogen (left panel) and the ion-imbalance method (right panel). The markers in the right panel distinguish between free water molecules and water coordinated to ions ($\rm{d_{O\text{-}Na^+} \leq \SI{3.3}{\angstrom}}$ or $\rm{d_{O\text{-}Cl^-} \leq \SI{3.9}{\angstrom}}$). The black line is the result of an equivalent, ion-free calculation at the potential of zero charge. (b) Representative snapshot of the interfacial water structure (1st and 2nd layers) and hydrogen-bond network (solid black lines). The formed rings are highlighted with colors.}
    \label{fig:hbonds}
\end{figure}

The orientational analysis in the previous section captures local, single-molecule information. To access the collective organization of interfacial water, we turn to the hydrogen-bond network, which provides a connectivity-level descriptor that captures collective organisation rather than single-molecule properties. We quantify it here through the donor/acceptor balance of each water layer and through ring statistics (elementary cycles in the H-bond graph, enumerated under periodic boundary conditions, see Computational Methods). The orientational analysis above already shows that the building units of the hydrogen-bond network at the interface vary with the applied charge. We analyzed the hydrogen-bond network within the first water bilayer using the geometric criterion introduced above ($d_{\mathrm{OO}} \leq \SI{3.5}{\angstrom}$ and $\alpha_{\mathrm{OOH}} \leq \SI{30}{\degree}$). The first water layer has a surplus of hydrogen-bond donors, while the second water layer compensates with a surplus of acceptors, as described by Le \textit{et al.}\cite{Le2018structure} This characteristic is recovered in both setups, as shown in Fig.~\ref{fig:hbonds}, and is consistent with water molecules being chemisorbed in a tilted $\mathrm{H \nearrow}$ configuration on the surface. This allows them to act as hydrogen-bond donors twice and as acceptors once within the first bilayer, and vice versa for $\mathrm{H \uparrow}$/$\mathrm{H \downarrow}$.

In bulk water, the numbers of hydrogen-bond donors and acceptors are roughly equal around the PZC. However, we find slightly more hydrogen bonds per water molecule in the IIM simulations. This difference is likely not caused solely by the net charge imbalance: the ion-free calculation (black line in Fig.~\ref{fig:hbonds}a) follows the same trend, and the 10--10 configuration with equal numbers of cations and anions (grey line) differs from the ion-free case. The offset is therefore consistent with the explicit-electrolyte environment in the IIM cell, including the different local coordination environments of Na$^+$ and Cl$^-$ and the resulting perturbation of hydrogen-bond donors and acceptors. In general, sodium- and chloride-coordinated water molecules form fewer water--water hydrogen bonds, while uncoordinated water molecules behave similarly to those in the ion-free simulation. The first-shell radial distribution functions for Na$^+$--O and Cl$^-$--H (Fig.~S14 in the SI, digitised from Ref.~\citenum{Kumar_thesis}) show that these ion solvation shells are essentially unchanged across the three IIM charge states, so the variation in interfacial hydrogen bonding above is driven by the ion populations rather than by changes in their solvation environment. This local hydrogen-bond perturbation is mirrored in the VDOS analysis below, where ion-coordinated water displays a higher O--H stretching frequency than uncoordinated water in the same bulk-like region.

We also investigated the hydrogen-bond network in the smaller PCH setup, where a larger charging window was explored (see Fig.~S2 in the SI). This analysis shows an increasing number of hydrogen bonds with positive electrode charge. For large biases, the numbers of hydrogen-bond donors and acceptors within the first water bilayer do not balance, leading to an overall orientation of water molecules in the middle of the cell. This effect may be related to the low screening of pure water.\cite{Huang2023} The increasing number of hydrogen bonds at the interface is reflected in the ring structure: at positive electrode charge the interface is composed of mostly five- and six-membered rings, while at negative electrode charge the water molecules form chains. A representative snapshot of the interfacial water structure at negative and positive electrode charges is shown in Fig.~\ref{fig:hbonds}b, and further details on the precise ring statistics are given in the SI. The ring statistics show similar results for the small and large PCH setups, whereas substantially more rings were found at the IIM interface, especially at the PZC (cf.\ Figs.~S3 and S4 in the SI). A similar reordering of the hydrogen-bond network was experimentally observed on gold.\cite{Velasco-Velez2014}

\section{Vibrational Analysis}

While atomistic calculations can readily access the topology of the hydrogen-bond network, this topology is not directly accessible to experiment at buried electrochemical interfaces. The most common indirect probe is vibrational spectroscopy of water,\cite{Kitadai2014} typically interpreted by comparison with atomistic simulations\cite{Wang2021,Zhu2025,Li2019,Zhang2020,Arvelos2025,Li2023,Li2025,Zhu2020,Zhu2024,Goldsmith2021,Tang2024,Li2025Elucidating,Li2022Unconventional,Li2022Hydrogen}. It remains an active area of research,\cite{Devlin_FD_2024} both because the interfacial signal is difficult to disentangle from the much larger bulk contribution and because the frequency-dependent dielectric response of metals can introduce significant artifacts into the measured spectra.\cite{Backus_JPCC_2012}

Despite these challenges, studies using diverse techniques such as attenuated-total-reflection surface-enhanced infrared absorption spectroscopy (ATR-SEIRAS),\cite{Gunasekaran_ChemSci_2023} shell-isolated nanoparticle-enhanced Raman spectroscopy (SHINERS),\cite{Wang2021} or plasmon-resonant vibrational sum-frequency generation (PR-VSFG)\cite{Wallentine_JPCA_2020,Dengh_ChemSci_2023} have reported vibrational spectra of interfacial water molecules as a function of applied voltage.

Layer-resolved vibrational spectra at the electrified Pt(111)/water interface were first computed as surface-sensitive VDOS (ss-VDOS) in Ref.~\citenum{Khatib2021nanoscale}, albeit only for the IIM charging method. Here, we employ the ss-VVCF formalism,\cite{ssVVCF} which yields Im\,$\chi^{(2)}$, whose amplitudes can take either sign at the same frequency and therefore encode molecular orientation, and 
extend that analysis 
in two important directions: (i) we compare vibrational spectra across both IIM and PCH, and
(ii) we disentangle the contributions of the four main orientations of interfacial water to the total signal.

\begin{figure}[!ht]
    \centering
    \begin{subfigure}[t]{\textwidth}
        \centering
        \includegraphics[width=\textwidth]{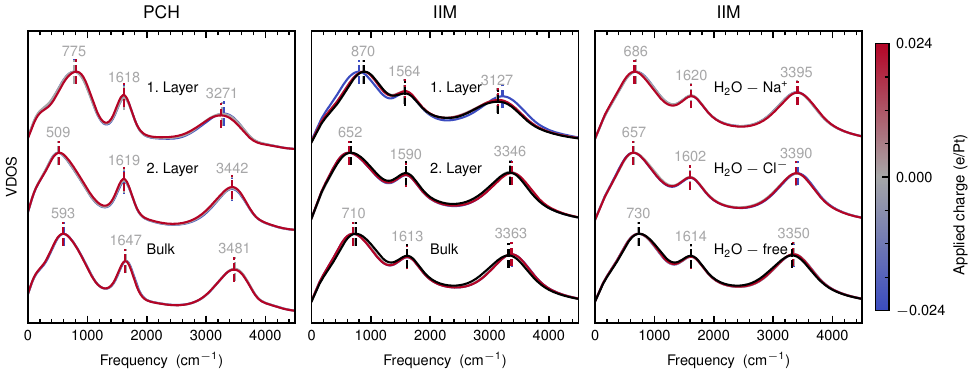}
        \caption{Spatially resolved vibrational density of states of water at applied charge. The left panel shows layer-resolved PCH spectra, the middle panel shows layer-resolved IIM spectra, and the right panel decomposes the IIM bulk-water spectra according to ion coordination. The spectra were calculated as the Fourier transform of the hydrogen--hydrogen velocity autocorrelation function.\cite{Le2018Theoretical} Gaussian smoothing ($\sigma \sim \SI{100}{\per\centi\metre}$) was applied. The librational band lies in the ${\approx}\,500$--$1000\,\mathrm{cm}^{-1}$ range, the H--O--H bending band near $1600\,\mathrm{cm}^{-1}$, and the O--H stretching band in the ${\approx}\,3000$--$3700\,\mathrm{cm}^{-1}$ range.}
        \label{fig:vdos}
    \end{subfigure}

    \vspace{0.5em}

    \begin{subfigure}[t]{0.95\textwidth}
        \centering
        \includegraphics[width=\linewidth]{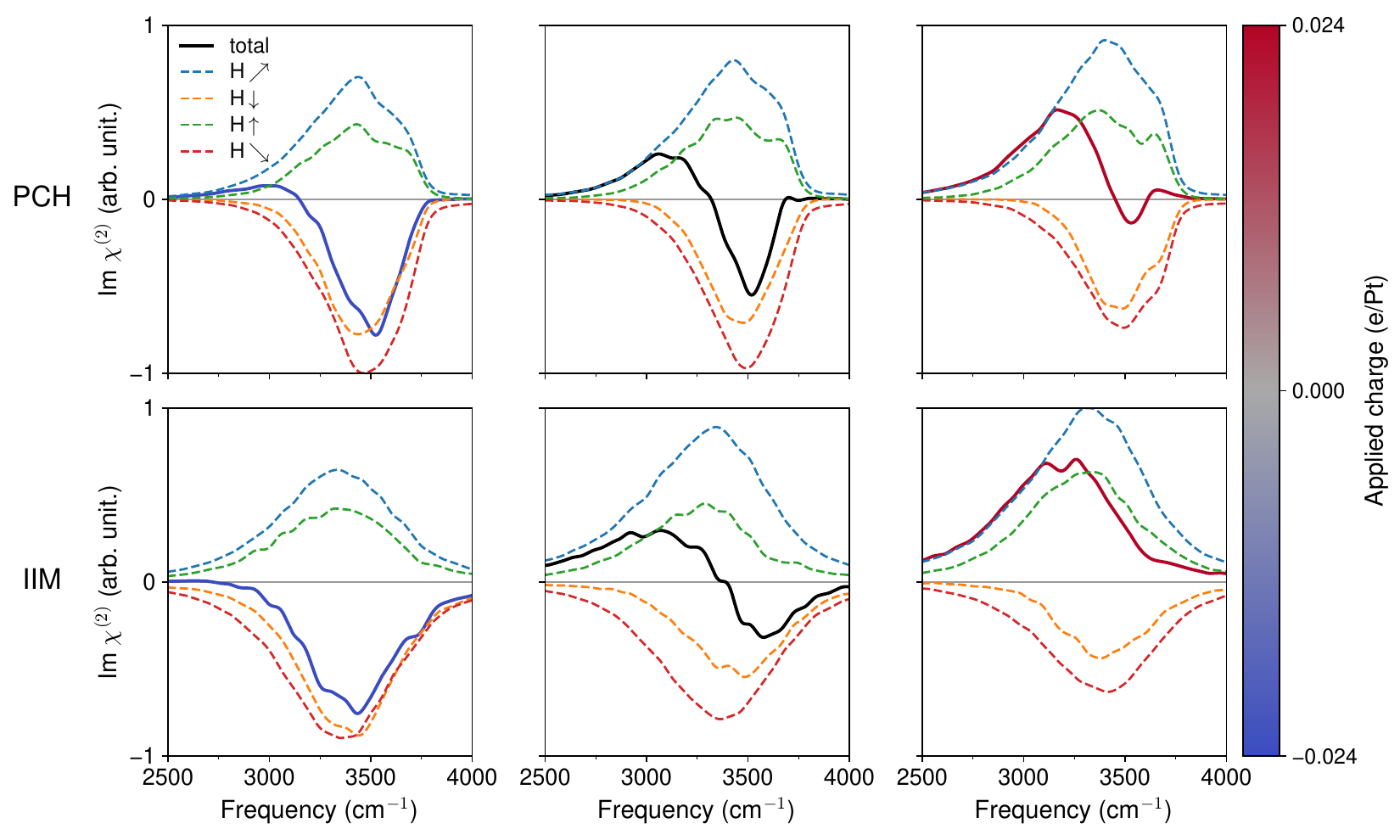}
        \caption{Predicted VSFG spectra of the Pt(111)/water interface at different applied charges for both the PCH (top) and IIM (bottom) methods. Dashed lines indicate the contribution from each orientational population. A positive (negative) signal corresponds to water molecules whose O--H bonds point away from (towards) the Pt substrate.}
        \label{fig:sfg}
    \end{subfigure}

    \caption{Vibrational analysis of interfacial water. (a) Spatially resolved VDOS and (b) predicted VSFG spectra at different applied charges.}
    \label{fig:vibrational_analysis}
\end{figure}

We begin by examining the layer-resolved VDOS, a surface-specific descriptor of O--H band positions for comparison with the ATR-SEIRAS and SHINERS measurements available at Pt and related Pt-group electrodes. Similar to Ref.~\citenum{Le2018Theoretical}, we calculated the hydrogen--hydrogen velocity autocorrelation functions and averaged the resulting spectra with respect to the position of the corresponding oxygen atom. In Fig.~\ref{fig:vibrational_analysis}a, we present the VDOS for the first and second water layers, as well as the bulk, as a function of the applied charge for both methods.

For the systems at the PZC, and across both methods, the bulk and second layers exhibit similar VDOS in the stretching and bending regions (the second-layer librational band is softened by 60--$80\,\mathrm{cm}^{-1}$, cf.\ Table~\ref{tab:water_structure_summary}), whereas the first layer displays a marked red shift in the high-frequency O--H stretching region (${\approx}\,3000$--$3700\,\mathrm{cm}^{-1}$) and a blue shift in the low-frequency librational region (${\approx}\,500$--$1000\,\mathrm{cm}^{-1}$) relative to the rest of the water. The layer contrast is the largest spectral feature: the first-layer O--H stretching band is red-shifted by roughly $200\,\mathrm{cm}^{-1}$ relative to bulk-like water (first two panels of Fig.~\ref{fig:vibrational_analysis}a). This shift is consistent with the flat, chemisorbed configuration and with the finite charge transfer between Pt and first-layer water identified in the Bader analysis. In contrast, the blue shift of the librational band indicates a stiffer rotational potential for chemisorbed water. The Pt--O bond of the flat-lying $\mathrm{H \nearrow}$ configuration directly hinders molecular rotation, and the resulting orientational constraint is accompanied by a more ordered hydrogen-bond environment; both effects act in the same direction and are consistent with reported DFT simulations.\cite{Le2018Theoretical} A similar stiffening effect has been observed at the water/air interface and attributed to the termination of the hydrogen-bond network.\cite{Tong_PCCP_2016} Detailed plots of the peak shifts are provided in the SI.

Charge-induced changes in the VDOS are particularly visible in the first layer but are comparatively smaller than the differences between layers. For both methods, the stretching band exhibits a red shift (blue shift) upon positive (negative) electrode charging, while the librational band shows the opposite trend. These shifts reflect a combination of interfacial electric-field effects, changes in hydrogen bonding, and charge transfer between Pt and chemisorbed water.\cite{Cassone_PCCP_2019} The fact that the response is concentrated in the first layer rather than spread uniformly across layers indicates that the shift originates primarily from the chemisorbed state of the interfacial water itself---its Pt--water bond, local geometry, and charge transfer---rather than from a bulk-uniform Stark effect acting on all water molecules equally. These charge-induced shifts agree with ATR-SEIRAS measurements at Pt electrodes, which report a red-shift of the O--H stretching band with increasingly anodic potential\cite{Zhu2020,Li2022Hydrogen,Li2025}; the spatial localisation of the shift in the chemisorbed first layer of the simulation is consistent with the surface-specific nature of that signal.

The IIM data further isolate electrolyte-specific spectral fingerprints (right panel of Fig.~\ref{fig:vibrational_analysis}a). In the bulk-like region, water molecules coordinated to Na$^+$ or Cl$^-$ show O--H stretching peaks tens of wavenumbers higher than uncoordinated water in the same IIM simulations (Fig.~S11). This high-frequency contribution is consistent with the reduced water--water hydrogen bonding of ion-coordinated molecules in Fig.~\ref{fig:hbonds}a and provides a microscopic assignment for electrolyte-induced spectral changes that would be absent in the ion-free PCH representation. The higher O--H stretching frequency we find for ion-coordinated water in the IIM matches the cation-dependent SEIRAS features reported in mixed-electrolyte experiments at Pt.\cite{Li2022Hydrogen,Li2025}

VSFG spectroscopy probes the second-order optical response, $\chi^{(2)}$, which vanishes in centrosymmetric media and is therefore inherently surface-specific.\cite{MoritaBook} With the sign convention used here, in which the surface normal points from Pt into the electrolyte, the sign of the amplitudes in the Im\,$\chi^{(2)}$ spectrum is linked to the orientation of water molecules within the O--H stretching response: positive (negative) amplitudes indicate that the O--H group points away from (toward) the Pt surface.

Because the polarizability-dipole correlation function converges only over much longer trajectories than are accessible to AIMD, we use the surface-specific velocity-velocity correlation function (ss-VVCF) formalism\cite{ssVVCF}, which builds Im\,$\chi^{(2)}$ from well-sampled atomic velocities and converges within the available simulation time.\cite{Morita_JPCB_2002,Litman_JPCL_2023b} 
In Fig.~\ref{fig:vibrational_analysis}b, we present the predicted VSFG spectra. At the PZC, the spectrum exhibits a broad bipolar lineshape, with a positive peak at approximately $3000~\mathrm{cm}^{-1}$ and a negative peak at approximately $3600~\mathrm{cm}^{-1}$, consistent with Ref.~\citenum{Khatib2021nanoscale}.

We further decompose the signal by orientation, using the four orientations ($\mathrm{H\nearrow}$, $\mathrm{H\uparrow}$, $\mathrm{H\downarrow}$, $\mathrm{H\searrow}$) defined above and in Fig.~\ref{fig:orientation}. This analysis shows that all four populations make significant contributions. More specifically, the $\mathrm{H \uparrow}$ and $\mathrm{H \nearrow}$ populations contribute positively to Im\,$\chi^{(2)}$, whereas the $\mathrm{H \downarrow}$ and $\mathrm{H \searrow}$ populations contribute negatively. The latter appear at slightly lower frequencies, suggesting that the hydrogen-bonding interaction between water and the metal is somewhat stronger than the water--water interactions. This interference gives rise to the bipolar lineshape.

At positive surface charge, the spectrum becomes predominantly positive, corresponding to water molecules whose O--H bonds point away from the interface. At negative surface charge, it becomes predominantly negative, corresponding to water molecules whose O--H bonds point toward the Pt substrate. The decomposition shows that this evolution reflects a transfer of population from the $\mathrm{H \uparrow}$ and $\mathrm{H \nearrow}$ orientations to the $\mathrm{H \downarrow}$ and $\mathrm{H \searrow}$ orientations as the surface charge becomes increasingly negative.

We note that the characteristic `free O--H' peak observed at the water/air interface and commonly associated with hydrophobic interfaces is absent here.\cite{Litman_NatChem_2024,Advincula_JACS_2026}

The two biasing schemes largely agree on Im\,$\chi^{(2)}$ at the PZC and on its dependence on surface charge, consistent with the agreement described in the previous sections. However, the PCH method produces somewhat narrower peaks, and small differences arise in the predicted frequencies of the individual populations. Most notably, at positive surface charge the PCH spectrum retains a small negative contribution, whereas at negative surface charge it retains a small positive contribution. Since these frequency differences are already present at the PZC, they do not originate from the biasing scheme itself. Both setups employ the same PBE exchange--correlation functional; we therefore attribute the residual offsets to the remaining differences in the computational protocols---thermostat and target temperature, plane-wave cutoff and basis-set completeness (\SI{300}{\rydberg} vs.\ \SI{400}{\rydberg}), and the inclusion of dispersion corrections in the PCH setup only---which are known to shift O--H stretching frequencies by a few tens of wavenumbers without altering the qualitative picture. That the two schemes nevertheless agree on the sign, the charge dependence, and the orientational decomposition of Im\,$\chi^{(2)}$ further highlights the robustness of these trends with respect to such methodological choices. Table~\ref{tab:water_structure_summary} summarises the structural, topological, and vibrational trends of both methods discussed in the preceding sections.

\begin{table}[!t]
\centering
\scriptsize
\setlength{\tabcolsep}{3pt}
\renewcommand{\arraystretch}{0.98}
\caption{Compact summary of charge- and electrolyte-dependent water-structure trends at electrified Pt(111), including gas-phase PBE H$_2$O monomer\cite{Santra_JCP_2009} and PBE ice-Ih harmonic\cite{Liu_JPCB_2016} reference frequencies. Positive/negative signs denote anodic/cathodic charging. All frequencies and frequency shifts are rounded to $10\,\mathrm{cm}^{-1}$. Frequency shifts are given relative to the respective bulk-water peak of each method, which leads to good quantitative agreement between the two computational setups.}
\label{tab:water_structure_summary}
\begin{tabularx}{\linewidth}{@{}>{\raggedright\arraybackslash}p{0.18\linewidth}%
                                >{\raggedright\arraybackslash}p{0.17\linewidth}%
                                >{\raggedright\arraybackslash}p{0.22\linewidth}%
                                Y@{}}
\doublerule
Property & PCH & IIM & Comment / Explanation \\
\doublerule

\sectionrow{Structure}
Avg. bilayer & same & same & charge-controlled \\
1st layer density & $+/-$: up/down & $+/-$: up/down & $+$ stabilizes chemisorbed H$_2$O \\
2nd layer density & $+/-$: down/up & $+/-$: down/up & $-$ stabilizes physisorbed H$_2$O \\
Orientational probabilities & linear vs. charge & linear vs. charge & local field response \\
\addlinespace[2pt]

\sectionrow{Topology}
\subsectionrow{\# H-bonds}
Avg. trend & $+/-$: up/down & $+/-$: up/down + offset & increase from chemisorbed-water anchoring \\
1st layer & 2.5--2.6 & 2.3--2.7 & donor-rich chemisorbed layer \\
2nd layer & 2.8--2.9 & 3.1--3.3 & acceptor-rich physisorbed layer \\
Bulk & 3.2--3.4 & all: 3.2--3.4; uncoord.: ${\approx}\,3.7$; ion-coord.: 2.3--2.6 & ions reduce water--water H-bonds \\
\subsectionrow{H-bond network}
Dominant motif & $-/+$: chains/rings & more rings & chain-to-ring transition likely from increased chemisorbed-water anchoring \\
\addlinespace[2pt]

\sectionrow{Vibrational properties}
\subsectionrow{Libration}
Ice Ih, PBE harmonic & 580--1030 cm$^{-1}$ & 580--1030 cm$^{-1}$ & Ref.~\citenum{Liu_JPCB_2016} \\
Bulk reference & 590 cm$^{-1}$ & 710 cm$^{-1}$ & computational setup differences \\
1st layer & +180 cm$^{-1}$ & +160 cm$^{-1}$ & stiffened rotation from chemisorption \\
2nd layer & $-$80 cm$^{-1}$ & $-$60 cm$^{-1}$ & softened rotation from reduced H-bonds \\
Bias shift & $+/-$: blue/red & $+/-$: blue/red & strongest in 1st layer \\
Ion coordination & absent & Na$^+$: $-$30; Cl$^-$: $-$50 cm$^{-1}$ & softening from reduced H-bonds vs. IIM bulk \\
\addlinespace[2pt]

\subsectionrow{H--O--H bend}
H$_2$O monomer, gas phase & 1600 cm$^{-1}$ & 1600 cm$^{-1}$ & Ref.~\citenum{Santra_JCP_2009} \\
Ice Ih, PBE harmonic & 1650--1710 cm$^{-1}$ & 1650--1710 cm$^{-1}$ & Ref.~\citenum{Liu_JPCB_2016} \\
Bulk reference & 1650 cm$^{-1}$ & 1610 cm$^{-1}$ & computational setup differences \\
1st layer & $-$30 cm$^{-1}$ & $-$50 cm$^{-1}$ & minor softening from reduced H-bonds \\
2nd layer & $-$30 cm$^{-1}$ & $-$20 cm$^{-1}$ & minor softening from reduced H-bonds \\
Bias shift & weak & weak & not main bias fingerprint \\
Ion coordination & absent & Na$^+$: +10; Cl$^-$: $-$10 cm$^{-1}$ & weak ion dependence \\
\addlinespace[2pt]

\subsectionrow{O--H stretch}
H$_2$O monomer, gas phase & 3700/3810 cm$^{-1}$ & 3700/3810 cm$^{-1}$ & Ref.~\citenum{Santra_JCP_2009} \\
Ice Ih, PBE harmonic & 3110--3360 cm$^{-1}$ & 3110--3360 cm$^{-1}$ & Ref.~\citenum{Liu_JPCB_2016} \\
Bulk reference & 3480 cm$^{-1}$ & 3360 cm$^{-1}$ & computational setup differences \\
1st layer & $-$210 cm$^{-1}$ & $-$240 cm$^{-1}$ & softened bond from chemisorption, opposite to libration \\
2nd layer & $-$40 cm$^{-1}$ & $-$20 cm$^{-1}$ & minor softening \\
Bias shift & $+/-$: red/blue & $+/-$: red/blue & red shift (predominantly 1st layer) by $+$ charging from increased H-bond network (\# H-bonds and rings) \\
Ion coordination & absent & Na$^+$: +30; Cl$^-$: +30 cm$^{-1}$ & minor stiffening from reduced H-bonds \\
\addlinespace[2pt]

\sectionrow{VSFG}
Im $\chi^{(2)}$ & $+/0/-$: pos./bipol./neg. & $+/0/-$: pos./bipol./neg. & preferential O--H orientation \\

\doublerule
\end{tabularx}
\end{table}

\section{Conclusions}

Unraveling the nanostructure and dynamics of electrified metal--water interfaces is essential for optimizing electrochemical technologies. In real electrochemical double layers, changes in electrode bias and changes in electrolyte ion distributions occur together, making it difficult to separate charge-controlled water response from ion-specific effects. In this work, we used two established constant-charge methods to model electrified interfaces within \textit{ab initio} molecular dynamics: the ion-imbalance method (IIM) and partially charged hydrogen atoms (PCH). By comparing results obtained via both biasing methods we identified which features of the interfacial water response are generic consequences of surface charge and which depend on the explicit electrolyte environment.

Surface charge provides a robust descriptor for the chemisorbed first layer and the integrated first water bilayer. Both methods yield a largely consistent structural picture when results are compared on the charge scale, despite differences in counter-charge representation, potential referencing, electrolyte composition, and cell size. This indicates that the average density and orientational response of the interfacial bilayer are primarily controlled by the electrode charge rather than by the microscopic representation of the counter-charge.
The hydrogen-bond network topology---donor/acceptor balance and elementary-ring statistics---emerges as a charge-sensitive, connectivity-level descriptor of the interfacial water response. In the extended PCH series, it maps a clean transition from chain-like motifs at negative electrode charge to ring-rich, ice-like motifs at positive electrode charge. The IIM data add complementary information on the electrolyte environment: explicit ions perturb water--water hydrogen bonding mainly through their local solvation shells and enhance ring populations near the PZC.

The vibrational analysis translates this structural picture into experimentally accessible vibrational fingerprints. The layer-resolved VDOS assigns the strongest O--H stretching perturbation to chemisorbed first-layer water, where the flat adsorption geometry and Pt--water charge transfer likely contribute strongly to the intramolecular O--H response. Explicit ions introduce an additional high-frequency contribution from ion-coordinated water, consistent with their reduced water--water hydrogen bonding. The computed VSFG spectra preserve the same qualitative two-band response in both charging setups, while the residual difference in the positive O--H stretching band points to enhanced sensitivity to physisorbed, electrolyte-facing water and identifies which parts of the double layer should be most visible to ATR-SEIRAS and VSFG experiments. The chain-like hydrogen-bond topology found at negative electrode charge is potentially relevant to the cathodic regime of hydrogen evolution, but its role in proton-transfer pathways remains a hypothesis for future reactive simulations.

Together, these analyses translate the charge-versus-ion distinction into concrete experimental targets. Charge-controlled bilayer reorganization should be visible as sign changes in Im\,$\chi^{(2)}$ at the physisorbed, electrolyte-facing layer; cation- or anion-specific SEIRAS features should trace the reduced water--water hydrogen bonding of ion-coordinated water in the bulk-like region; and the hydrogen-bond ring statistics provide a connectivity-level fingerprint of the surface-charge state. These observable-level fingerprints define a concrete route by which future experiments can separate charge-controlled and ion-specific contributions to interfacial water structure at Pt(111).

\begin{acknowledgement}
This work was supported by the Engineering and Physical Sciences Research Council (grant EP/P033555/1). We also gratefully acknowledge the use of the High-Performance Computers at Imperial College London, provided by the Imperial College Research Computing Service (\url{https://doi.org/10.14469/hpc/2232}). This research also used the ARCHER2 UK National Supercomputing Service (\url{https://www.archer2.ac.uk}) via our membership of the UK's HEC Materials Chemistry Consortium, which is funded by EPSRC (EP/R029431 and EP/X035859). Finally, we would like to acknowledge the Thomas Young Centre under grant number TYC-101.

\end{acknowledgement}

\begin{suppinfo}

Additional plots of average water density and orientation (Fig.~S1), hydrogen-bond network analysis for the extended PCH charge range (Fig.~S2), ring statistics for PCH and IIM (Figs.~S3--S4), hydrogen-bond metrics (Figs.~S5--S8), vibrational spectra and peak shifts (Figs.~S9--S12), a Bader charge table (Table~S1), per-molecule Bader charges over the extended PCH charge range (Fig.~S13), and ion-coordination radial distribution functions for the IIM cells (Fig.~S14).

\end{suppinfo}

\bibliography{sample}

\end{document}


\section{Average Water Density and Orientation}

\begin{figure}[!p]
    \centering
    \begin{subfigure}[t]{0.68\textwidth}
        \centering
        \includegraphics[width=\linewidth]{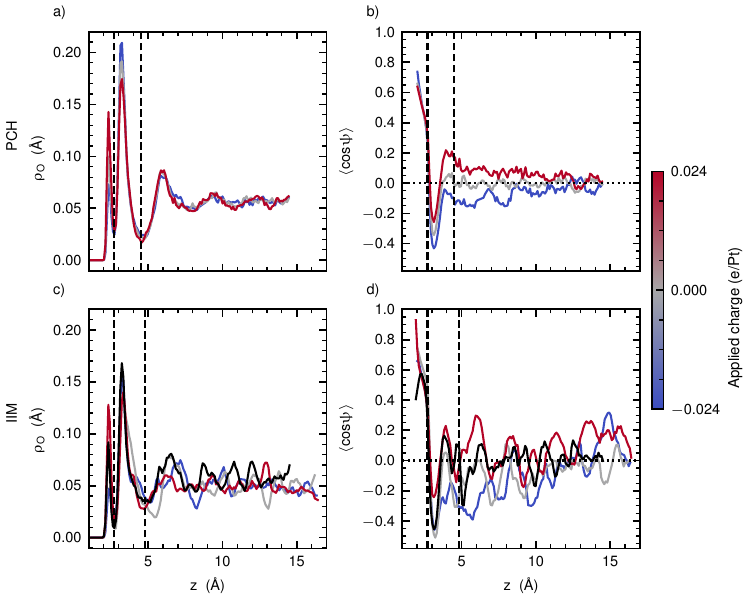}
        \caption{Water oxygen density and water orientation, represented by the bisector angle $\psi$, along the surface normal.}
    \end{subfigure}

    \vspace{0.5em}

    \begin{subfigure}[t]{0.68\textwidth}
        \centering
        \includegraphics[width=\linewidth]{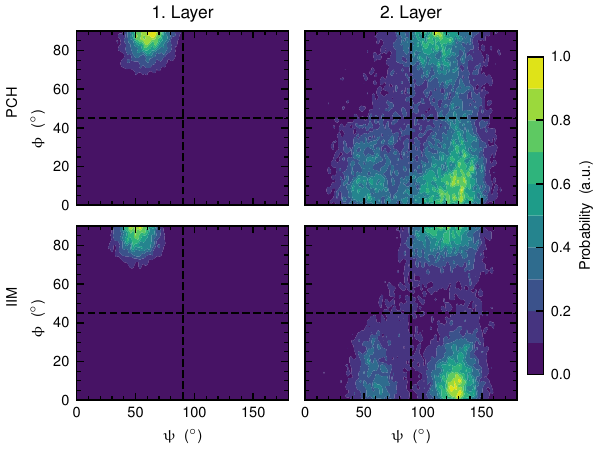}
        \caption{Layer-resolved distribution of cartwheel angle $\psi$ and propeller angle $\phi$ at the potential of zero charge.}
    \end{subfigure}

    \caption{Average density and orientational response of water at Pt(111). The upper panels in each plot correspond to PCH and the lower panels to IIM. The layer-resolved angular distributions distinguish first-layer chemisorbed water from second-layer physisorbed water and support the layer definitions used throughout the main text.}
    \label{fig:S_density_orientation}
\end{figure}

\clearpage
\section{Hydrogen-Bond Network and Ring Statistics}

\begin{figure}[!htbp]
    \centering
    \includegraphics[width=\textwidth]{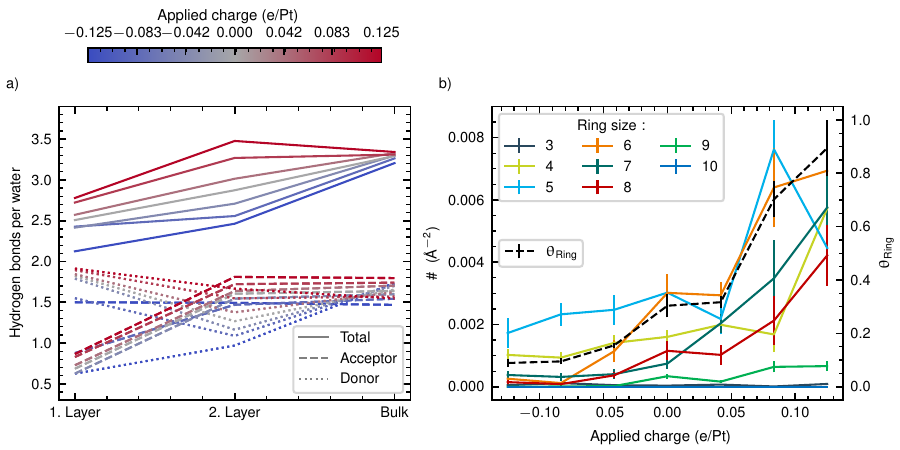}
    \caption{Hydrogen-bond donors and acceptors per water molecule and ring statistics for the smaller PCH setup over the extended charge range. The analysis shows the transition from chain-like networks at negative electrode charge to ring-rich, ice-like networks at positive electrode charge and indicates that the trend persists beyond the narrower charge range used for the method comparison in the main text.}
    \label{fig:S_hbond_pch_extended}
\end{figure}

\begin{figure}[!htbp]
    \centering
    \includegraphics[width=\textwidth]{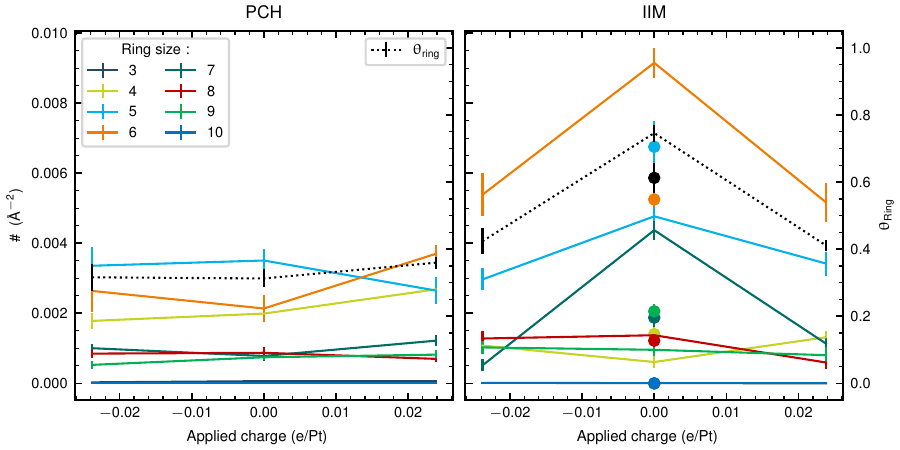}
    \caption{Ring formation in the interfacial water bilayer for PCH and IIM. Colored solid lines indicate the coverages of rings of different sizes, and the black dotted line denotes the total ring coverage. The ion-free calculation at the potential of zero charge is marked by circles. The comparison highlights that IIM exhibits a substantially stronger ring population around the PZC, whereas the narrow-range PCH data show a weaker variation. The extended PCH data in Fig.~S4 show the broader transition from chain-like networks at negative electrode charge to ring-rich networks at positive electrode charge.}
    \label{fig:S_rings_large}
\end{figure}

\begin{figure}[!htbp]
    \centering
    \includegraphics[width=\textwidth]{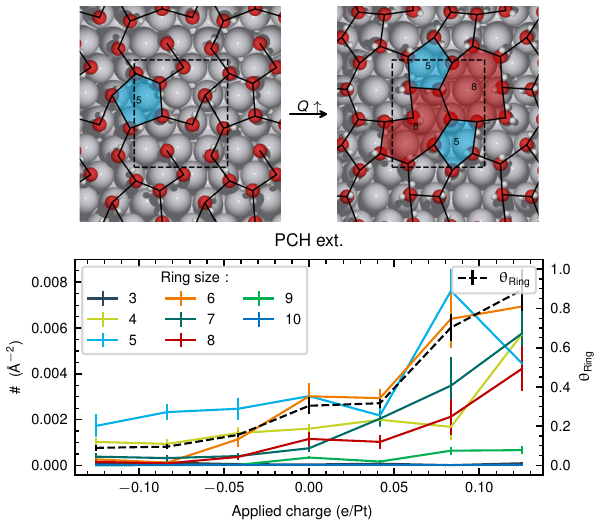}
    \caption{Ring statistics and representative snapshots for the smaller PCH setup over the extended charge range. The snapshots illustrate the structural transition from chain-like networks under negative electrode charge to ring-rich networks under positive electrode charge.}
    \label{fig:S_rings_extended}
\end{figure}

\clearpage
\section{Hydrogen-Bond Metrics}

\begin{figure}[!htbp]
    \centering
    \includegraphics[width=\textwidth]{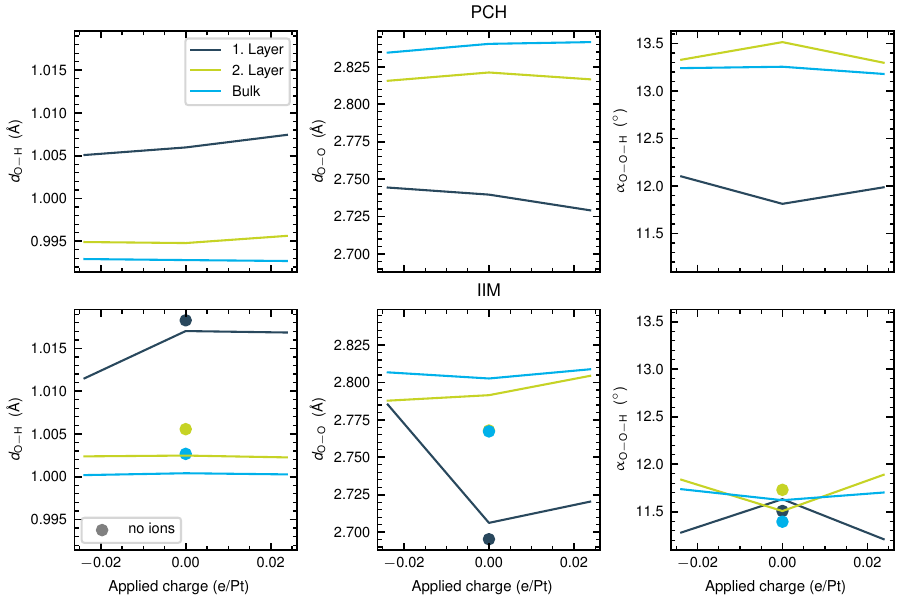}
    \caption{Average hydrogen-bond metrics as a function of applied charge for PCH and IIM. The panels show the intramolecular O--H distance, the intermolecular O--O distance, and the hydrogen-bond angle $\alpha_{\mathrm{O-O-H}}$. These averages show that changes in network topology are not accompanied by large discontinuous changes in the local hydrogen-bond geometry.}
    \label{fig:S_hb_metrics_ave}
\end{figure}

\begin{figure}[!htbp]
    \centering
    \includegraphics[width=\textwidth]{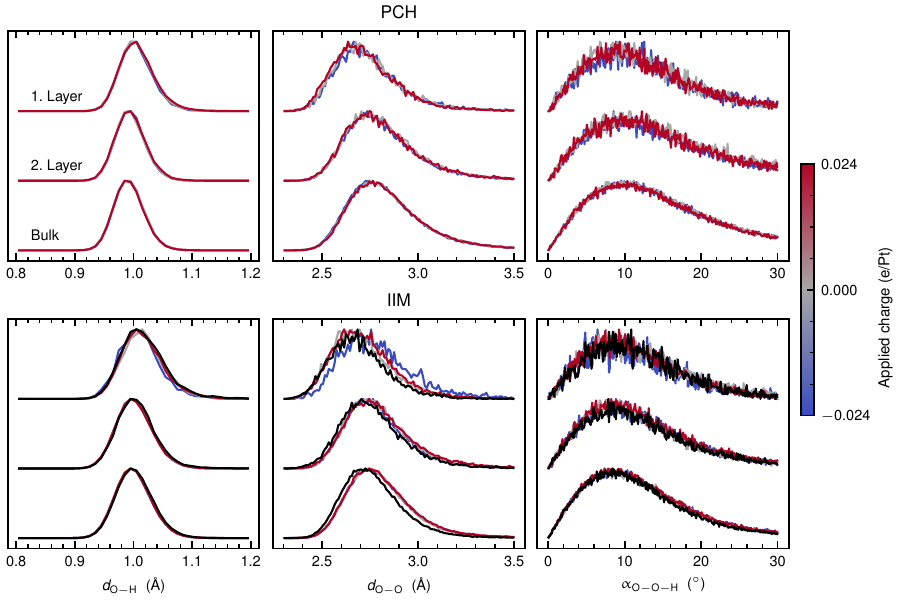}
    \caption{Distributions of hydrogen-bond metrics for PCH and IIM. The panels show the intramolecular O--H distance, the intermolecular O--O distance, and the hydrogen-bond angle $\alpha_{\mathrm{O-O-H}}$ resolved by water layer and applied charge. The distributions complement the average metrics by showing how the first and second layers contribute differently to the interfacial hydrogen-bond environment.}
    \label{fig:S_hb_metrics_dist}
\end{figure}

\begin{figure}[!htbp]
    \centering
    \includegraphics[width=\textwidth]{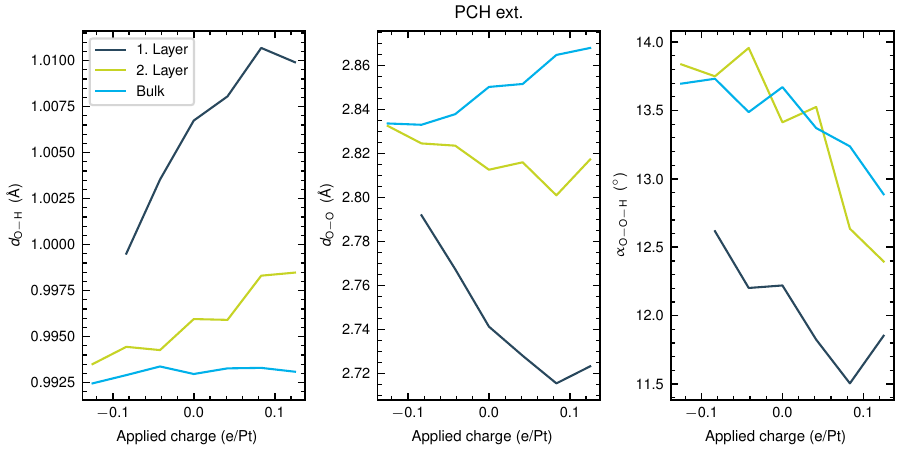}
    \caption{Average hydrogen-bond metrics for the smaller PCH setup over the extended charge range. The extended range shows that the hydrogen-bond response evolves continuously with charge and does not rely on the specific charge points used in the direct PCH/IIM comparison.}
    \label{fig:S_hb_metrics_ave_extended}
\end{figure}

\begin{figure}[!htbp]
    \centering
    \includegraphics[width=\textwidth]{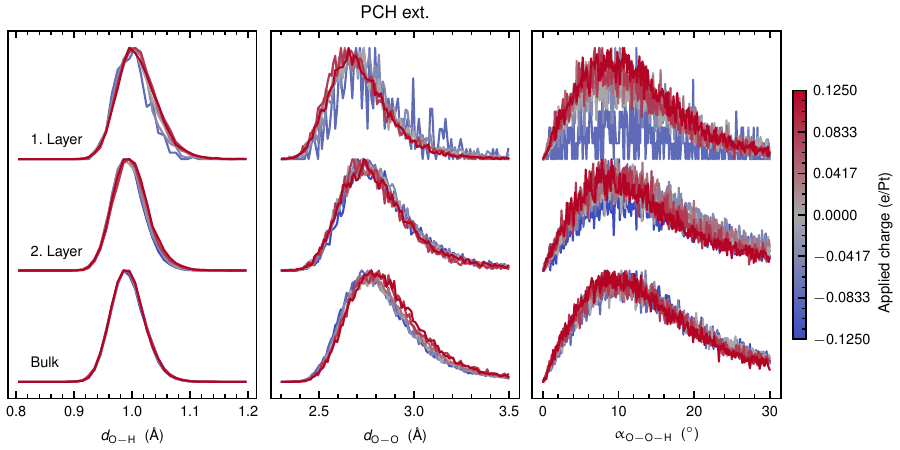}
    \caption{Distributions of hydrogen-bond metrics for the smaller PCH setup over the extended charge range. These distributions provide the microscopic basis for the average trends in Fig.~S7 and for the network-topology changes discussed in the main text.}
    \label{fig:S_hb_metrics_dist_extended}
\end{figure}

\clearpage
\section{Vibrational Spectra and Peak Shifts}

The figures in this section provide VDOS/VACF information resolved by water layer and, for IIM, by ion coordination. The figures in this section provide vibrational power spectra (VDOS/VACF) resolved by water layer and, for IIM, by ion coordination. Combined with the layer-resolved orientation distributions of Fig.~\ref{fig:S_density_orientation}, they account for the VSFG band assignment discussed in the main text.

\begin{figure}[!htbp]
    \centering
    \includegraphics[width=\textwidth]{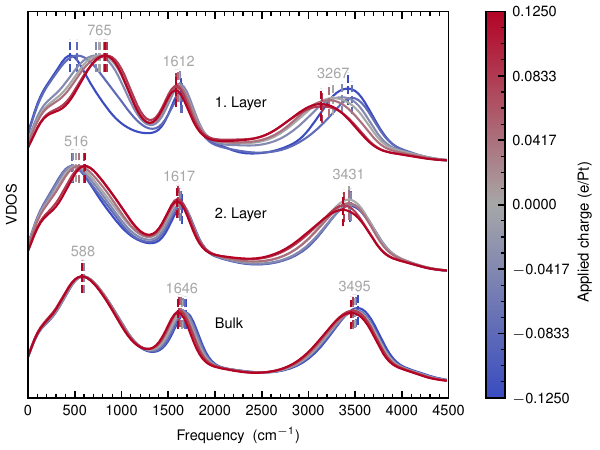}
    \caption{Layer-resolved vibrational density of states for the smaller PCH setup over the extended charge range. Spectra were obtained from hydrogen--hydrogen velocity autocorrelation functions and resolved according to the oxygen position of the corresponding water molecule. The plot shows that the largest spectral contrast is between chemisorbed first-layer water and more bulk-like water, while charge modulates this layer-dependent response.}
    \label{fig:S_vacf_extended}
\end{figure}

\begin{figure}[!htbp]
    \centering
    \includegraphics[width=\textwidth]{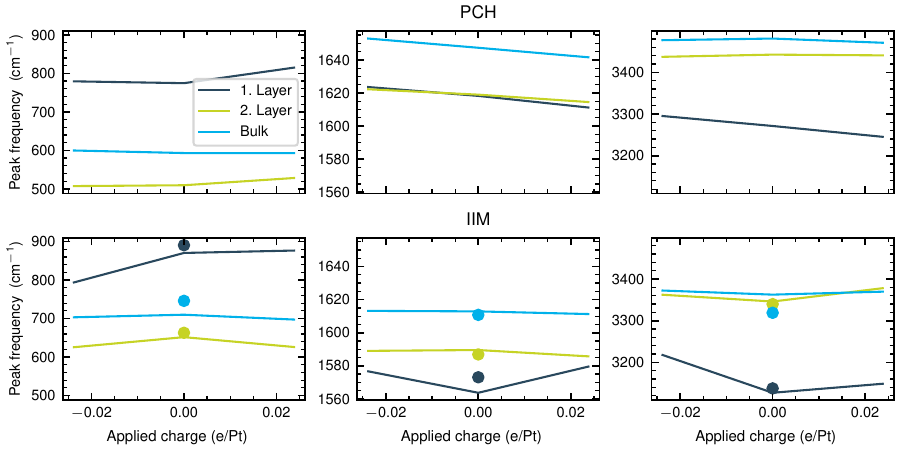}
    \caption{Peak frequencies of the librational, bending, and stretching regions of the VDOS as a function of applied charge for PCH and IIM. The ion-free calculation is marked by circles. The charge-dependent O--H stretching shifts are most pronounced in the first layer, consistent with the stronger perturbation of chemisorbed water at Pt(111).}
    \label{fig:S_peak_shift}
\end{figure}

\begin{figure}[!htbp]
    \centering
    \includegraphics[width=\textwidth]{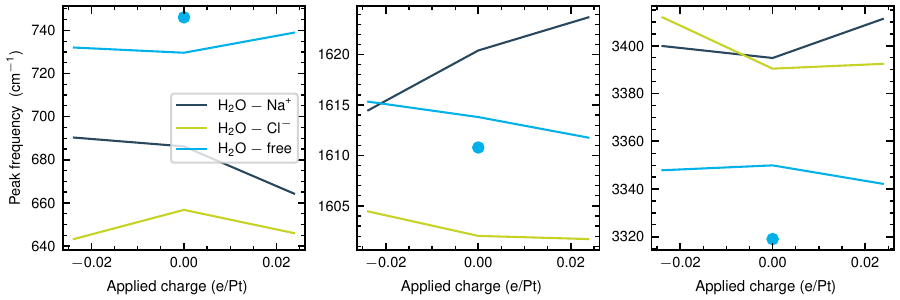}
    \caption{Peak frequencies of the VDOS for bulk water in the IIM setup, separated into water coordinated to ions and uncoordinated water molecules. Ion coordination was defined using $d_{\mathrm{O-Na^+}} \leq \SI{3.3}{\angstrom}$ or $d_{\mathrm{O-Cl^-}} \leq \SI{3.9}{\angstrom}$. Ion-coordinated water exhibits higher O--H stretching frequencies than uncoordinated water in the same IIM environment, providing a spectral signature of explicit-electrolyte solvation.}
    \label{fig:S_peak_shift_ions}
\end{figure}

\begin{figure}[!htbp]
    \centering
    \includegraphics[width=\textwidth]{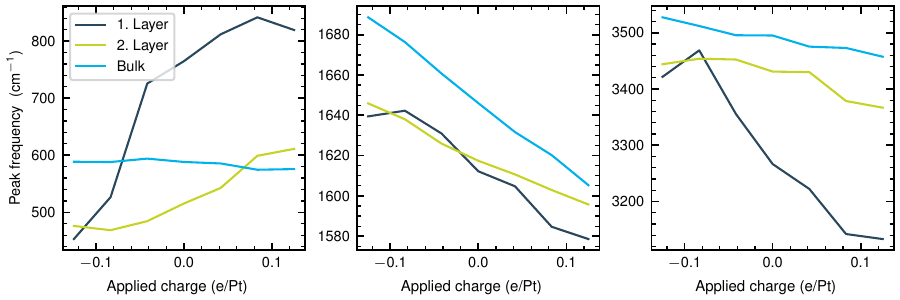}
    \caption{Peak frequencies of the VDOS for the smaller PCH setup over the extended charge range. The extended series shows that the first-layer stretching and librational frequencies evolve continuously with electrode charge.}
    \label{fig:S_peak_shift_extended}
\end{figure}

\clearpage
\section{Bader Charge Analysis}

\begin{table}[!htbp]
    \centering
    \small
    \caption{Bader charge analysis for the electrode and water layers. IIM values are reproduced from Ref.~\citenum{Khatib2021nanoscale}. All charges are reported in units of $|e|$ per simulation cell. The water-layer columns list total layer charges, not per-molecule charges; per-molecule normalization therefore requires the corresponding number of water molecules in each layer. The ``sum'' column is defined as the sum of the subsurface Pt, surface Pt, first water layer, and second water layer contributions. Missing bulk entries indicate quantities that were not used in the corresponding sum.}
    \label{tab:S_bader}
    \begin{tabular}{lrrrrrrrr}
        \hline
        Method & Nominal charge & \multicolumn{3}{c}{Electrode charge} & \multicolumn{3}{c}{Water charge} & Sum \\
        \cline{3-5} \cline{6-8}
         & & Total & Subsurface & Surface & Layer 1 & Layer 2 & Bulk & \\
        \hline
        IIM & 0 (no ions) & -0.56 & 1.65 & -2.21 & 0.62 & -0.09 & -- & -0.03 \\
        IIM & -1 & -1.10 & 1.70 & -2.80 & 0.29 & -0.23 & -- & -1.04 \\
        IIM & 0 & -0.61 & 1.68 & -2.29 & 0.49 & -0.16 & -- & -0.28 \\
        IIM & +1 & -0.03 & 1.67 & -1.70 & 0.77 & -0.08 & -- & 0.66 \\
        PCH & -1 (small cell) & -1.09 & 1.43 & -2.52 & 0.52 & -0.34 & -0.13 & -0.91 \\
        PCH & 0 (small cell) & -0.58 & 1.41 & -1.98 & 0.70 & -0.17 & -0.01 & -0.05 \\
        PCH & +1 (small cell) & -0.04 & 1.40 & -1.44 & 0.88 & 0.01 & 0.12 & 0.85 \\
        \hline
    \end{tabular}
\end{table}

\begin{figure}[!htbp]
    \centering
    \includegraphics[width=0.7\textwidth]{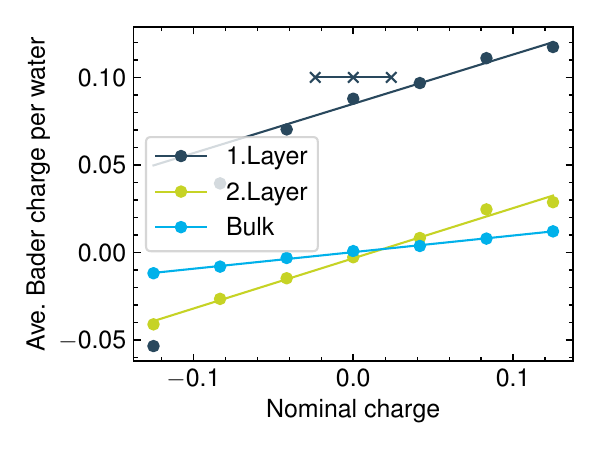}
    \caption{Average Bader charge per water molecule as a function of nominal applied charge for the smaller PCH setup over the extended charge range. The dark blue line shows first-layer water, the yellow line second-layer water, and the light blue line bulk water. The cross markers indicate the corresponding IIM values at the three comparison charge states. In the narrow comparison range used in the main text, the first-layer per-molecule charge is approximately $0.07$--$0.11\,|e|$ for PCH and nearly constant at ${\sim}\,0.10\,|e|$ for IIM. Over the full extended PCH range, the first-layer charge varies more strongly (from ${\sim}\,-0.05$ to ${\sim}\,0.12\,|e|$), while second-layer and bulk water remain close to zero throughout.}
    \label{fig:S_bader_permolecule}
\end{figure}

\clearpage
\clearpage
\section{Ion Coordination in IIM Cells}

\begin{figure}[!htbp]
    \centering
    \includegraphics[width=\textwidth]{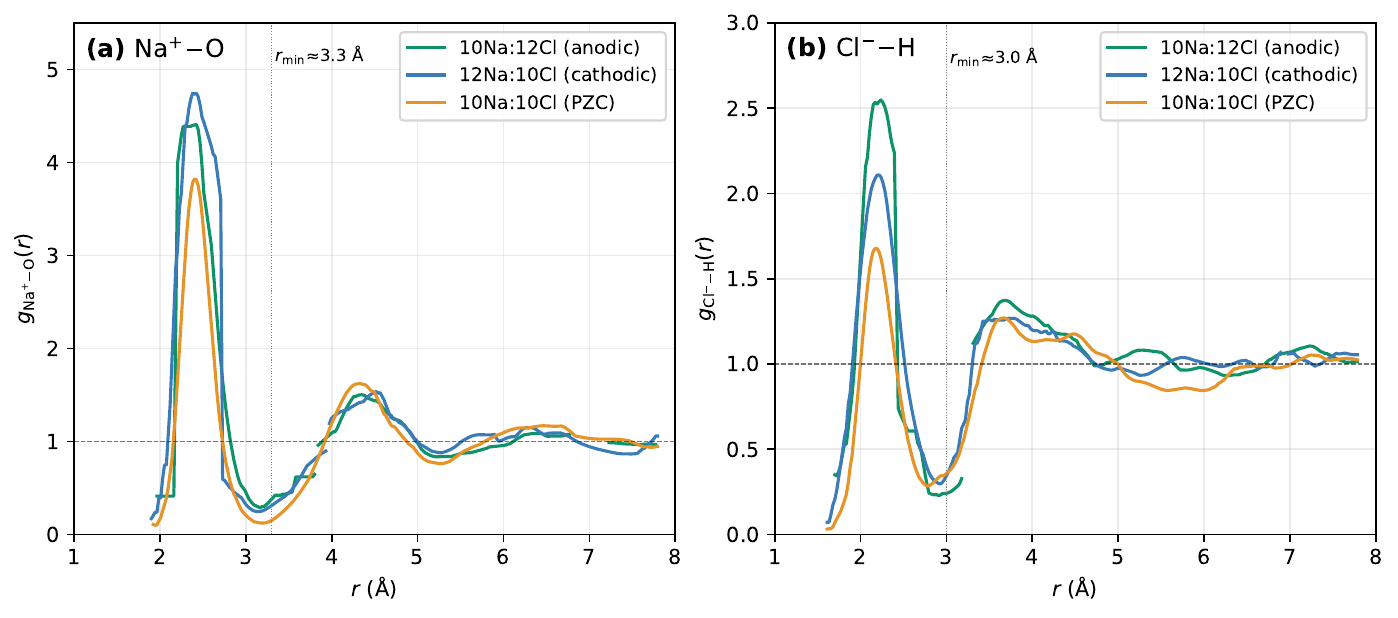}
    \caption{Radial distribution functions $g(r)$ of (a) Na$^+$--O and (b) Cl$^-$--H for the three IIM cells used in this work, labelled in the main-text Na:Cl convention: 10Na:12Cl (anion excess, anodic), 10Na:10Cl (neutral, PZC), and 12Na:10Cl (cation excess, cathodic). The horizontal dashed line marks $g(r{\to}\infty)=1$ and the vertical dotted lines mark the first-minimum cutoffs used in the main-text hydrogen-bond analysis: $r_{\min}\!\approx\!\SI{3.3}{\angstrom}$ for Na$^+$--O (panel a) and $r_{\min}\!\approx\!\SI{3.0}{\angstrom}$ for Cl$^-$--H (panel b), corresponding to a Cl$^-$--O cutoff of ${\sim}\,\SI{3.9}{\angstrom}$. The three cells give very similar first-shell shape and width, indicating that the ion solvation structure itself is largely insensitive to the surface-charge state of the electrode. Curves were digitised from Figs.~C-2 and C-3 of Ref.~\citenum{Kumar_thesis} (Appendix~C), with the Ag(111) reference curve in the original figures omitted here for clarity, and renormalised to the standard $g(r{\to}\infty)=1$ convention by dividing by the median asymptotic value over the range $r=6\text{--}\SI{7.8}{\angstrom}$.}
    \label{fig:S_coord_rdf}
\end{figure}

\bibliography{sample}